\documentclass[%
 reprint,
 amsmath,amssymb,
 aps,
pra,
]{revtex4-1}

\usepackage{color}
\usepackage[table]{xcolor}
\usepackage{graphicx}
\usepackage{dcolumn}
\usepackage{bm}
\usepackage{lipsum}
\usepackage{placeins}
\usepackage{verbatim}

\definecolor{lightlightgrey}{gray}{0.92}

\newenvironment{exer*}
  {\ex}
  {\endex}

\def\be{\begin{equation}}
\def\ee{\end{equation}}
\def\bea{\begin{eqnarray}}
\def\eea{\end{eqnarray}}

    \newcommand{\opdagger}[2]{\mbox{$\hat{#1}_{#2}^{\dagger}$}}
    \newcommand{\op}[2]{\mbox{$\hat{#1}_{#2}$}}

\usepackage{xspace} 

\newcommand{\hairsp}{\hspace{1pt}} 
\newcommand{\ie}{\textit{i.\hairsp{}e.}\xspace} 

\begin{document}
\title{Release-free silicon-on-insulator cavity optomechanics}

\author{Christopher J. Sarabalis}
\email{sicamor@stanford.edu}
\author{Yanni D. Dahmani}
\author{Rishi N. Patel}
\author{Jeff T. Hill}
\author{Amir H. Safavi-Naeini}
\email{safavi@stanford.edu}
\affiliation{%
Department of Applied Physics and Ginzton Laboratory, Stanford
University, Stanford, California 94305, USA
}%

\begin{abstract}

We demonstrate optically coupled nanomechanical resonators fabricated on silicon-on-insulator. Silicon fin waveguides are used to control the dispersion of mechanical waves and engineer localized resonances by modulation of the fin properties. A photonic crystal cavity is designed to localize laser light near the fin and the mechanical motion is read out and modified by radiation pressure back-action. We expect devices and systems made from similar structures to enable co-integration of signal transduction and processing capabilities via electronic, photonic, and phononic degrees freedom in a single scalable platform.

\end{abstract}

\date{ \today}

\maketitle

\section{Introduction}
The ability to confine and guide photons with low loss in thin films of silicon on silicon oxide has made silicon-on-insulator (SOI) a leading platform for photonic circuits~\cite{Soref2006}. Integrating mechanical devices into the same platform would greatly enhance the capabilities of the silicon photonics toolbox. Unfortunately, due to silicon's mechanical properties, mechanical waves, or phonons, are not guided in the device layer and hence oxide release processes are needed to suspend the silicon device and confine mechanical motion~\cite{Li2008,Eichenfield2009,Eggleton2013}. This greatly limits co-integration of electronic, photonic, and phononic elements and prevents us from making silicon phononic circuits that place mechanical and optical waves on the same footing in the SOI platform. Moreover, releasing the silicon devices limits their ability to dissipate heat greatly complicating cryogenic optomechanical experiments~\cite{Meenehan2014}. In this work, we demonstrate a silicon-on-insulator nanomechanical device that uses fins to confine motion to the device layer without requiring a release process.  Incorporating these fins into the first SOI optomechanical device, we use laser light to transduce their motion.  The large optomechanical coupling in these structures allows us to observe thermal Brownian motion of the fin, as well as the optomechanical spring effect~\cite{Aspelmeyer2014b}. Our demonstration opens a route to new silicon optomechanical mass and force sensors, acousto-optic modulators, optical and microwave filters, as well as hybrid electronic-photonic-phononic structures that take advantage of the coupled dynamics in a semiconductor to obtain novel functionality.

In this work we demonstrate the first SOI optomechanical devices by fashioning optical and mechanical resonators from fin waveguides. We begin in section~\ref{sec:fin_mech} by describing the physics of the fin mechanical waveguides and resonators. We then outline in section~\ref{sec:optics} how optical resonators can be patterned into the same device layer and describe the optomechanical experiment. Finally in section~\ref{sec:optomechanical_coupling}, we present measurements of the thermal mechanical Brownian motion in a structure with an engineered multimode spectrum and deduce the optomechanical coupling rates from the optical spring effect.

\section{Confining mechanical waves on SOI}
\label{sec:fin_mech}

\begin{figure}[htbp]
    \centering
    \includegraphics[width=\linewidth]{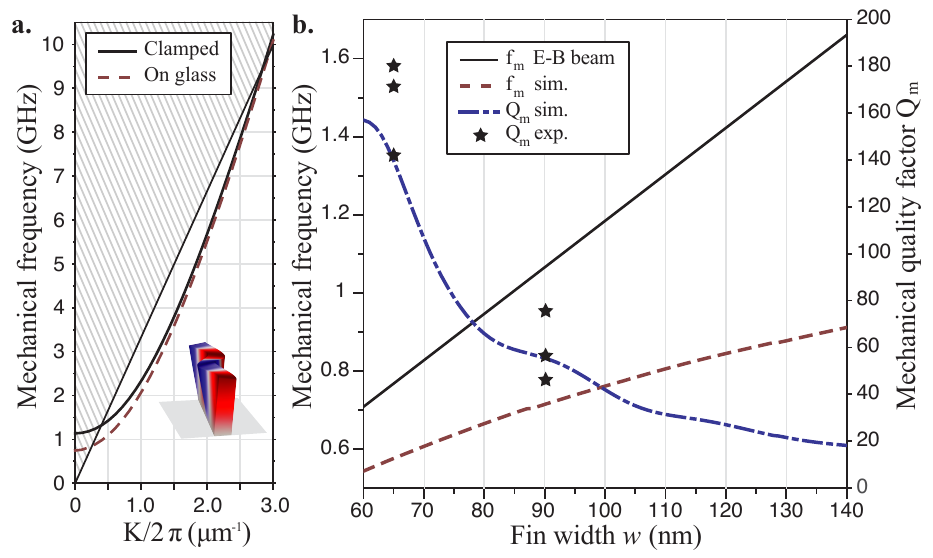}
    \caption{\textbf{a.} The dispersion diagram for guided waves in silicon fins with dimensions $h=340~\text{nm}$, $w=100~\text{nm}$, on an oxide substrate. \textbf{b.} Mechanical frequencies for the E-B model of a fin (solid line), and a fin on glass (dashed line) are compared. The E-B model starts to break down for small values of $h/w$ and there is a deviation from linearity in the full solution. Additionally, the fins on oxide generally have lower frequency due to the ``softer'' clamping boundary condition. Simulated radiation limited quality factors (dot-dashed line) for the fins are overlayed with $Q_\text{m}$ measurements (stars).  Corresponding solutions are marked with a red circle on \textbf{a.} and \textbf{b.}.}
    \label{fig:mechPhys}
\end{figure}

At optical frequencies, the index contrast between silicon ($n_\textrm{Si}=3.48$) and silicon oxide ($n_\textrm{Ox}=1.44$) enables optical waveguiding in SOI via total internal reflection. In contrast to optical waves, mechanical waves or phonons propagate more quickly in silicon than in silicon oxide. This makes obtaining total internal reflection of mechanical waves in silicon clad with silicon oxide impossible. One approach to obtaining confinement that has been used in piezoelectric bulk acoustic wave devices~\cite{Hashimoto2009} and more recently in silicon resonant body transistors~\cite{Bahr2015} is to pattern a mechanical reflector directly into the substrate to prevent the mechanical waves from leaking away. A different approach  is to sufficiently reduce the mechanical wave propagation phase velocity in the silicon layer so that the leakage is eliminated. Despite inconvenient material parameters, phonons can be confined in this way by leveraging the role that geometry and free boundaries play in their  dispersion. A key observation is that tall, thin devices are more compliant than short, wide ones. We find that fin waveguides, such as that shown in Figure~\ref{fig:mechPhys} can  guide phonons on SOI. Similar mechanical waveguides on the surface of crystals were studied several decades ago~\cite{Ash1969,Mason1971}, and more recently, we have shown theoretically that fins on SOI can guide both photons and phonons and mediate large interactions between them~\cite{Sarabalis2016}.

A simple approximation for the fundamental resonance frequency of a silicon fin with width $w$, etched into a silicon device layer of thickness $h$ is obtained from Euler-Bernoulli (E-B) beam theory~\cite{Cleland2003}:
\bea
    \Omega \approx  \sqrt{\frac{E}{\rho}} \frac{w}{h^2}. 
\eea
The Young's modulus, $E$, and density, $\rho$, of silicon are taken to be $165~\text{GPa}$ and $2330~\text{kg/m}^3$ respectively and we assume an isotropic medium. This motion has no longitudinal variation and represents a $K=0$ point for the dispersion of phonons as shown in Figure~\ref{fig:mechPhys}a. For larger $K$-vectors, the dispersion of this band can cause it to have a phase velocity in the longitudinal direction smaller than any other mechanical wave of the system, which leads to lossless propagation of waves in silicon on oxide~\cite{Sarabalis2016}. This is represented by the crossing of the band from the hatched region to the unhatched, ``protected'' region in Figure~\ref{fig:mechPhys}a. In this work, we focus our attention on the unprotected region in this diagram. Considering that our devices are limited by fabrication to aspect ratios of $h/w < 10$, and the quality factor due to phonon leakage from a cantilever is known to be approximately $Q_\text{m} \approx C(h/w)^3$ where $C$ is a factor on the order of unity~\cite{Yasumura2000}, we expect the mechanical quality factors of our structures to be dominated by radiation of acoustic waves into the substrate. The effect of this mechanical radiation into surface and bulk acoustic waves can be modeled numerically and is plotted for different geometric parameters in Figure~\ref{fig:mechPhys}b. Finite-element method simulations in COMSOL~\cite{COMSOL5} of the equations of elasticity were performed to compute the dependence of mechanical frequency and $Q_\text{m}$ on fin width for $h=340~\text{nm}$ silicon-on-insulator. As shown in Figure~\ref{fig:mechPhys}b, increasing the width of the beam causes an increase in the frequency, in qualitative agreement with the approximate E-B theory, as well as a reduction in the mechanical quality factor. 

The fin geometry enables transverse confinement of motion in silicon-on-insulator waveguides~\cite{Sarabalis2016}. It is important to also longitudinally confine the mechanical motion for many devices of interest. To do so we break the translational symmetry by smoothly varying the fin's width to make a curved fin as shown in Figure~\ref{fig:optics}a. This smoothly modulates the cut-off frequency (the $K=0$ mode) of the fin waveguide causing it to support, in the thinned part, modes with frequencies below the cut-off frequency at the thicker edges of the fin. The mechanical mode profiles of the first few symmetric modes of such a structure are shown in Figure~\ref{fig:optics}b with exaggerated displacements. Since the modulation is smooth, we expect the $Q_\text{m}$ of these localized resonances to be approximately the same as the $Q_\text{m}$ of the guided waves. This is borne out in measurements described below and summarized in Figure~\ref{fig:mechPhys}b.

\section{The optics of curved fin cavities}\label{sec:optics}

\begin{figure*}[htbp]
    \centering
    \includegraphics[width=1.0\textwidth]{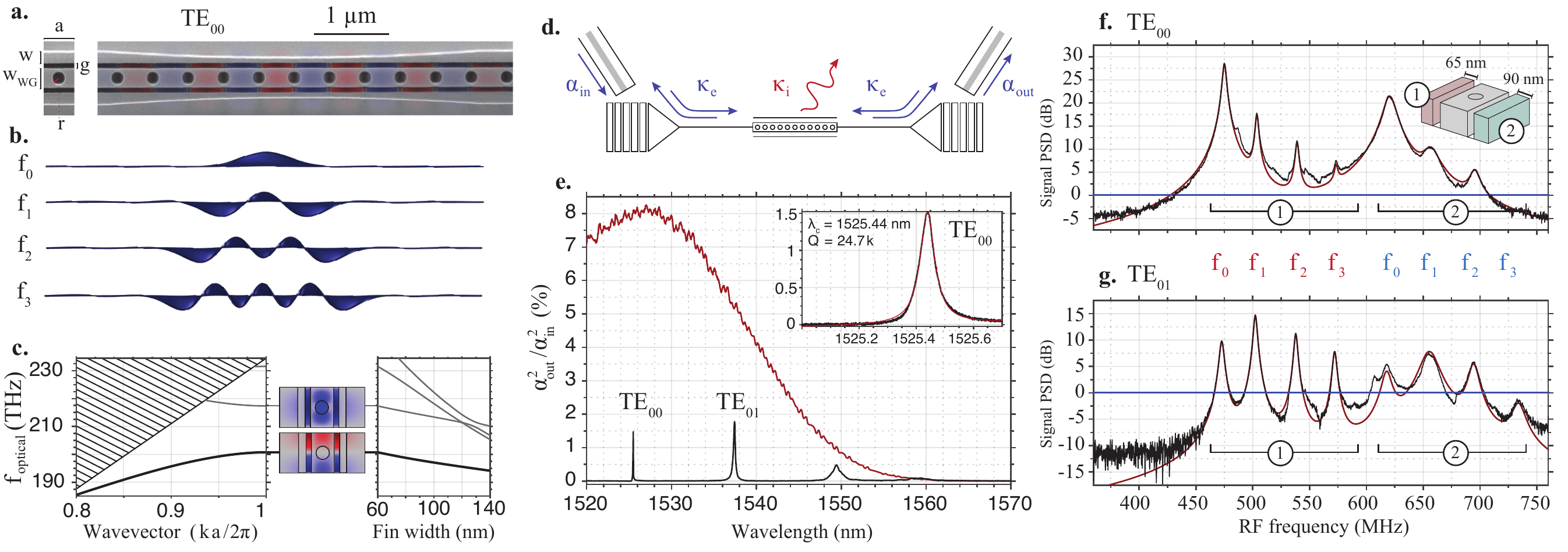}
    \caption{\textbf{a.} SEM of the fabricated structure composed of two fins surrounding a wide beam and forming a photonic crystal cavity. The unit cell of the photonic crystal is shown with the geometrtic parameters defined. \textbf{b.} Modulating the width of the fins leads to localized mechanical resonances with modes labeled $f_0$ to $f_3$ (these are only the even modes; odd modes are optically dark). \textbf{c.} The TE optical bands of a symmetric fin cavity unit cell have a 17~THz bandgap. The same variation that leads to trapped phonons (\textbf{b.}) causes optical resonances to be trapped in the central region. The transverse component of the electric field for one such mode ($\textrm{TE}_{00}$) is overlayed in the SEM in \text{a}. \textbf{d.} Measurement scheme: cleaved fibers are aligned to TE grating couplers.  Optical transmission spectra are recorded on an output channel and intensity fluctuations induced by the mechanics are read out on reflection. \textbf{e.}  An optical transmission spectrum for the fin cavity reported here (black) is plotted alongside the transmission of a through waveguide (red). In the inset, a narrower scan of the $\textrm{TE}_{00}$ mode is shown. The detected RF power spectral density for the laser tuned to the slope of the cavity resonance transmission for optical modes $\textrm{TE}_{00}$ and $\textrm{TE}_{01}$ are shown in \textbf{f.} and \textbf{g.}, respectively.}
    \label{fig:optics}
\end{figure*}

In order to optically read out the motion, the curved fins are incorporated into a photonic crystal cavity. Figure~\ref{fig:optics}a is a scanning electron micrograph (SEM) of a curved fin cavity consisting of a photonic crystal waveguide with adjacent curved fins. The unit cell of such a photonic crystal is characterized by a lattice constant $a$, optical waveguide width $w_\text{WG}$, circular hole radius $r$, fin width $w$, and a gap $g$. These geometric parameters are outlined in Figure~\ref{fig:optics}a. Photonic waveguides that are symmetric with respect to transverse reflections have modes that are antisymmetric (``TE'') or symmetric (``TM'') under a transverse reflection. Figure~\ref{fig:optics}a shows the TE optical bands for this unit cell. We focus on TE bands since these waves  exhibit a larger bandgap, though we have also made similar structures with TM guided waves. The TE resonances have a 17~THz bandgap that varies with $w$. Conveniently, the same perturbation that confines mechanical modes, \ie reducing the width of the fins, increases the frequency of the fundamental TE band as shown Figure~\ref{fig:optics}c, pushing the photonic crystal band edge into the bandgap and leading to confined optical resonances. 

We measure transmission of a laser light through a photonic crystal with the curved fin defect. Laser light from an external cavity diode laser (Santec TSL-550-A) is swept over a range of wavelengths (1520 nm - 1570 nm), and the resulting transmitted field is detected on a photodiode. Devices with curved fins give rise to the TE$_{00}$, TE$_{01}$, and TE$_{02}$ resonances, which appear as peaks in the transmission spectrum in Figure~\ref{fig:optics}e. Fits to this spectrum yield $(Q,\kappa_\textrm{e}/2\pi,\kappa_\textrm{i}/2\pi)$ of (25k, 2.8~GHz, 2.6~GHz) and (6.5k, 12~GHz, 5.1~GHz) for the first two resonances. Optical radiation loss rates as computed using FEM simulations in COMSOL~\cite{COMSOL5} are expected to be an order of magnitude lower than measured values suggesting that intrinsic optical losses are dominated by disorder and surface roughness. The total transmitted power from one fiber into the other is on the order of $1.5\%$ of the input power, with a large part of the insertion loss occurring at the grating couplers, measured separately to have $17 \pm 1\%$ transmission.

Due to the very small amount of mechanical energy in the substrate, the two curved fins of symmetric fin cavities have essentially degenerate mechanical spectra that are only different due to inhomogeneity or disorder. In order to remove this degeneracy so that we can more clearly resolve the modes, the two fins are fabricated with different widths. By making the center of the fins 65 and 90~nm wide, and having the widths  increasing parabolically in both directions by 30 nm over 7.5 $\mu$m we cause a shift of 145~MHz between the fundamental frequencies of the two fins. This perturbation however has consequences for the optical spectrum as it breaks the transverse symmetry plane and induces scattering between the quasi-TE and quasi-TM modes. For our cavity, simulations show an asymmetry-induced loss rate of $\kappa_\textrm{TM} = 2\pi\times$370~MHz from the nearly TE mode into propagating TM waves.

We expect improvements in the design of the photonic crystals to give us access to the sideband resolved regime, as well as to higher order and possibly symmetry protected mechanical modes in these structures. Nonetheless, the design presented has a number of very desirable features. The mechanical and optical design problems are largely decoupled. The fins can be designed first to engineer a mechanical response of interest. The photonic waveguide then offers independent degrees of freedom for engineering the optical modes. For the cavity measured and discussed below, $a$, $w$, and $r$ are not varied along the cavity.

\section{Optomechanical Coupling}\label{sec:optomechanical_coupling}

The mechanical motion of the fin's resonances, described by modal displacements $x_k$, cause fluctuations in the $n^\textrm{th}$ optical resonance frequency $\omega_{\textrm{opt},n}$. Here, index $n$ refers to either of the $\textrm{TE}_{00}$ or $\textrm{TE}_{01}$ modes. The coupling is described perturbatively by the relation $\omega_{\textrm{opt},n}(x_0, x_1, \ldots) = \omega_{\textrm{opt},n}(0) + \sum_k g_{\textrm{OM},nk} x_k$. The coupling rate for each mode pair $g_{\textrm{OM},nk}$ contains boundary and photoelastic contributions. Expressions for these contributions can be found in Ref.~\cite{Safavi-Naeini2014a} in terms of the mode profiles. The resulting optomechanical interaction Hamiltonian is given by
\bea
H_{\text{int}}= \sum_{k,n} \hbar g_{0,nk} (\opdagger{b}{k} + \op{b}{k}) \opdagger{a}{n} \op{a}{n},
\eea
where $\op{b}{k}$ and $\op{a}{n}$ are the annihilation operators for the $k^\textrm{th}$ mechanical mode and the $n^\textrm{th}$  optical mode respectively, and $g_{0,nk}$ are the respective single-photon optomechanical coupling rates. 

Optomechanical coupling to thermally excited mechanical degrees of freedom causes the intensity of light reflected off the cavity to fluctuate. We detect these fluctuations by first amplifying the light coming back from the cavity in reflection using an erbium-doped fiber amplifier (EDFA, Fiberprime EDFA-C-26G-S11) and sending the amplifier output to a photodetector (Optilab PD-40-M). The resulting photocurrent is sent to a spectrum analyzer (Rohde \& Schwarz FSW26). Representative detected RF spectra of the thermal Brownian motion of the fins detected on the TE$_{00}$ and TE$_{01}$ resonances are shown in Figure~\ref{fig:optics}f and g, respectively. For each spectrum, an RF spectral density $S_{VV}(\omega)$ is taken at a given detuning. In addition a far off-resonant ($~\text{nm}$) spectrum is recorded $S_{VV,\textrm{BG}}(\omega)$. The plotted signal is $S_{VV}(\omega)/S_{VV,\textrm{BG}}(\omega) - 1$, so the noise level is simply $1$ and the $y$-axis can be interpreted as the signal-to-noise ratio. The noise level is dominated by amplified spontaneous emission from the EDFA.

\begin{figure}[ht]
    \centering
    \includegraphics[width=0.5\textwidth]{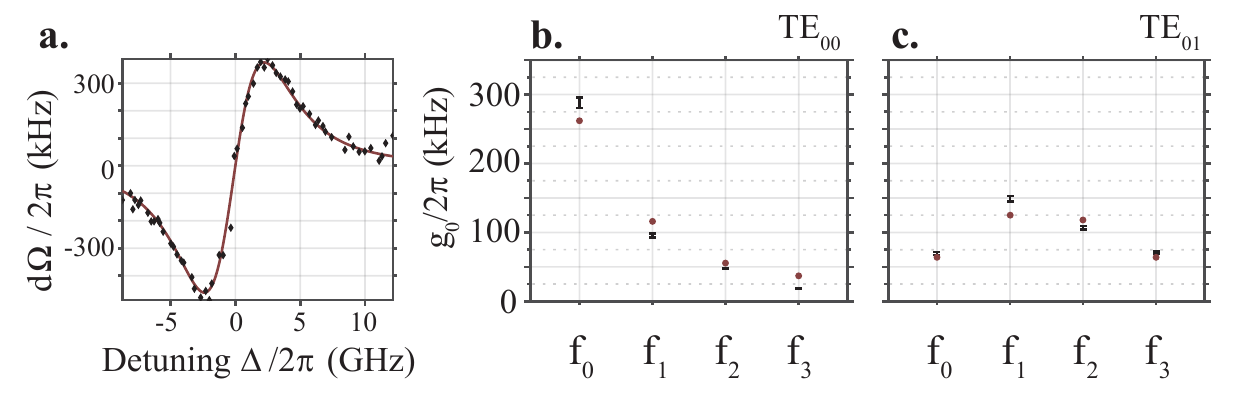}
    \caption{\textbf{a.} The optical mechanical spring effect is fit to determine the optomechanical coupling rate for TE$_{00}$ and $f_0$ modes. \textbf{b.} The resulting $g_{0,nk}$ between TE$_{00}$ the four mechanical modes of the 65 nm fin are fit (as described in the text) and seen to agree well with simulated coupling rates. \textbf{c.} Same as \textbf{b.} but for optical mode TE$_{01}$.}
    \label{fig:mechanics}
\end{figure}

To understand the mechanical and optomechanical response of the system and compare to theory, we extract the mechanical quality factors $Q_{\text{m},k}$ and the optomechanical coupling rates $g_{0,nk}$ for each mechanical mode $k$ from the measured RF spectra. The mechanical $Q$s are easily inferred by fitting the RF spectra and are plotted in Figure~\ref{fig:mechPhys}b for the two  fins. To extract the $g_{0,nk}$, we take advantage of the dependence of radiation pressure back-action effects on the coupling rate and the intracavity photon number, which can be independently calibrated to good precision. This approach has the advantage of not requiring precise knowledge of the gain of the optical and electronic amplifiers and detectors. Laser light modifies the dynamics of the mechanical resonator and can either stiffen ($\Delta > 0$) or soften ($\Delta < 0$) the mechanical mode depending on the detuning $\Delta = \omega_\textrm{L} - \omega_{\text{opt},0}$. This change in the mechanical frequency, denoted as $\textrm{d}\Omega_k$, is known as the \emph{optical spring effect}~\cite{Aspelmeyer2014b} and can be expressed as
\begin{equation}
    \textrm{d}\Omega_k = \frac{2 (g_{0,nk}|\alpha|)^2\Delta}{\left(\Delta - \Omega_k\right)^2 + \frac{\kappa_\textrm{t}^2}{4}},
    \label{eq:OSE}
\end{equation}
where $\kappa_\textrm{t} = 2 \kappa_\textrm{e} + \kappa_\textrm{i}$ is the total optical loss rate, and $|\alpha|^2$ is the intracavity photon number in the $n^\textrm{th}$ optical mode. For a fixed optical input power, we record and fit mechanical spectra to obtain the  mechanical frequency for a range of detunings $\Delta$. The detuning itself is verified by measuring the cavity transmission to cancel out thermal shifts of the cavity frequency. The resulting shift in the mechanical frequency and fits to Equation~\ref{eq:OSE} are shown in Figure~\ref{fig:mechanics}a. This procedure, repeated at different incident optical powers, is used to find the coupling rate of the TE$_{00}$ and f$_0$, TE$_{00}$ and f$_1$, and TE$_{01}$ and f$_1$ modes resulting in couping rates $g_{0,nk}/2\pi$ of $290 \pm 10$, $95 \pm 3$, and $150 \pm 5$~kHz, respectively. The systematic errors are larger than statistical errors, and are primarily due to the uncertainty in the photon flux incident on the cavity. Since the spring effect is proportional to $g_{0,nk}^2|\alpha|^2$, uncertainty in the intracavity photon number is directly propagated onto the coupling rate estimate. 
For modes with smaller optomechanical coupling, the spring effect is an unreliable means of determining the coupling. We compare contributions to the area under the power spectral density from each of these modes, to that of a mode where $g_{0,nk}$ has been determined using the spring effect. Assuming all modes for a single spectrum are at the same temperature, the power contributed by each peak scales as $g_{0,nk}^2/\Omega_k^2$, allowing precise determination of $g_{0,nk}$ even for modes with small optomechanical coupling. 

The resulting estimates of the coupling rates  for the two optical modes are compared to simulations of the interaction performed in COMSOL~\cite{COMSOL5} and are found to be in excellent qualitative and quantitative agreement as shown in Figure~\ref{fig:mechanics}b and c.

\section{Conclusions}\label{sec:conclusions}

We have demonstrated the first fully release-free silicon-on-insulator optomechanical devices. Further work is required to make structures that access the protected mechanical modes. We expect these mechanical modes to have very large  quality factors despite being fully connected to an oxide substrate. Since our devices do not require special release steps, they are completely compatible with silicon photonic foundry processes, can be produced at scale, and integrated easily with on-chip electronics and photonics. Our demonstration is a step towards silicon \emph{phononic} systems with the potential to combine optical, mechanical, and electronic functionality in an integrated platform.


\section*{Funding Information}
This work was supported by NSF ECCS-1509107, ONR MURI QOMAND, and start-up funds from Stanford University. ASN is supported by the Terman and Hellman Fellowships. RNP is supported by the NSF Graduate Research Fellowships Program. YDD is supported by the  Stanford UAR Major Grants program.

\section*{Acknowledgments}

The authors thank Raphael Van Laer for critical reading of the manuscript and valuable input.








\end{document}